\documentclass[preprint,showpacs,preprintnumbers,amsmath,amssymb,aps]{revtex4}
\usepackage{epsfig}
\usepackage{psfrag}
\usepackage{graphicx}
\usepackage{color,graphics}

\begin{document}

\title{New nonlinear structures in a degenerate one-dimensional electron gas}

\author{S. Ghosh}
%\email{titomend@ist.utl.pt}
\affiliation{Department of Applied Mathematics, University of Calcutta,
    92, Acharya Prafulla Chandra Road, Kolkata-700 009, India}

\author{N. Chakrabarti}
\affiliation{Saha Institute of Nuclear Physics, 1/AF Bidhannagar Calcutta - 700 064, India}

\author{F. Haas \footnote{Present address: Instituto de F\'isica, Universidade Federal do Rio Grande do Sul, CEP 91501-970, Porto Alegre RS, Brasil}}
%\email{ferhaas@fisica.ufpr.pr}
\affiliation{Departamento de F\'isica, Universidade Federal do Paran\'a, CEP 81531-990, Curitiba, PR, Brasil}

\begin{abstract}
The collective dynamics of nonlinear electron waves in an
one-dimensional degenerate electron gas is treated using the Lagrangian
fluid approach. A new class of solutions with a
nontrivial space and time dependence is derived. Both
analytical and numerical results demonstrate the formation of
stable, breather-like modes, provided certain conditions are meet. For 
large amplitude of the initial density perturbation, a catastrophic collapse of
the plasma density is predicted, even in the presence of the quantum statistical pressure and
quantum diffraction dispersive effects. The results are useful for the understanding of the properties of
general nonlinear structures in dense plasmas.
\end{abstract}


\maketitle

\newcommand\emc{E=mc^{2}}
\newcommand{\ep}{\epsilon}
\newcommand{\gr}{\nabla}
\newcommand{\ti}{\times}
\newcommand{\iy}{\infty}
\newcommand{\im}{\rightarrow}
\newcommand{\fr}{\frac}
\newcommand{\Pl}{\partial}
\newcommand{\pa}{\parallel}
\newcommand{\pe}{\perp}
\newcommand{\ts}{\textstyle}
\newcommand{\bee}{\begin{equation}}
\newcommand{\ene}{\end{equation}}
\newcommand{\beea}{\begin{eqnarray}}
\newcommand{\enea}{\end{eqnarray}}
\newcommand{\fder}[2]{\frac{{\ts d \/ #1}}{{\ts d\/ #2}}}
\newcommand{\fpar}[2]{\frac{{\ts \Pl \/ #1}}{{\ts \Pl \/ #2}}}
\newcommand{\nder}[3]{\frac{{\ts d^{#1} \/ #2}}{{\ts d \/ #3^{#1}}}}
\newcommand{\npar}[3]{\frac{{\ts \Pl^{#1} \/ #2}}{{\ts \Pl \/ #3^{#1}}}}
\newcommand{\sech}{\mbox{sech$\,$}}

\section{Introduction}

The collective dynamics of electrons in one-dimensional (1D) nanostructures opens up many new interesting avenues for
research because of its application in modern nanotechnology
\cite{tg04,weis}. A 1D geometry can be created
from a two-dimensional electron gas at a cleaved or gated
heterojunction \cite{ag91,oa05}. The dynamics
is qualitatively different when the charge carriers are confined in
1D channels or wires, because then the electrostatic interaction is
stronger than in higher dimensions. In fact,
new collective excitations arises in 1D. This is the case e.g. for
plasmons in pyroelectric-semiconductor composites \cite{dmitriev}.
The response of electrons confined to one-degree of freedom
exhibits exotic phenomena like spin-charge
separation and the emergence of correlated-electron insulators
\cite{vd10,kin}. The spin-charge separation phenomenon
supports electron charge density and spin density waves. In a different context, nowadays we have an increasing ability
to control plasma-fabricated metal-based nanostructures into 1D plasmonic devices \cite{rider}. The consideration of
quantum effects in such ultra-small systems is becoming increasingly unavoidable. In addition, the target normal sheath acceleration \cite{passoni} arising
from high intensity laser-matter interaction gives an example of dense 1D plasmas, specially with the development of coherent brilliant X-ray radiation sources \cite{thiele}.
However, the associated
self-consistent electronic states are highly
nonlinear in nature \cite{ai09,gb10}.
Therefore, the sensible analytic modeling of strongly localized structures at nanoscale is of
considerable interest.

In a multi-dimensional electron gas, numerical simulations predict
solitonic structures, when the electron wave packet interacts with
the metal surface. These stable structures are formed due to the
balance between the electron density-metal surface interactions
induced nonlinearity and the wave dispersion
\cite{sb05}. Periodic nonlinear structures appear in magnetized quantum plasmas too \cite{harris}.

In this Letter, we present for the first time a nonlinear
quasi-collapsing analytical solution in the
realm of 1D electron waves described by a quantum hydrodynamic model. The
electron density can become singular at a finite time due to the strong
effectiveness of the Coulomb interaction in 1D. Therefore, under certain circumstances to be
specified later, the Fermi
pressure and tunneling effects are shown to be unable to prevent the electron
density collapsing. %It is argued that the results of this investigation are relevant in general, e.g. for the transport of information in ultra-small 1D electronic structures. 
For definiteness, the focus will be in metallic nanostructures, although the results apply to a much broader class of systems.

%The transport phenomena of most of the liquids are described by
%the hydrodynamical theory when the spatial length scale of the
%system $(\lambda)$ is long compared to the particle-particle
%mean-free path $({\it l})$. In metallic nanostructures, at
%strictly zero temperature, all electrons have energies below the
%Fermi temperature $T_F$ and no transition takes place. As a result
%electron-electron collisions are not possible. Even at moderate
%temperature (room temperature) the electron-electron collision
%rate $\sim 10^{-10} s$ is much larger than the typical
%collisionless (plasmon oscillation) time scale $\sim 10^{-16} s$
%\cite{maha}. This makes the quantum (electron-electron) coupling
%parameter less than unity and $\lambda\gg {\it l}$. This allows to
%describe the electron dynamics at nanoscale hydrodynamically
%\cite{maha,mb97,domps,nc08}. Even in the strongly correlated
%electron systems, hydrodynamic approach can also be realized
%\cite{aa11}. Here, we consider the dynamics of electrons (short
%time scale $\sim \omega_p^{-1}$ inverse of plasmon frequency) and
%therefore ions are assumed to be immobile with charge neutrality
%background. ]
In metallic nanostructures, the thermodynamic temperature is orders of
magnitude smaller than the Fermi temperature $T_F$, hence the equation of
state for a degenerate electron gas is indicated. Moreover, the electron-electron
collision time scale $\sim 10^{-10} s$ is typically much larger than the plasmon
oscillation time scale $\sim 10^{-16} s$, justifying the neglect of collision
terms \cite{haas, maha} in a first approximation. Nevertheless, the gross features of the electron
dynamics in e.g. metal clusters can be described using fluid models, as found from
comparison with numerical simulation of more expensive models involving e.g. the Boltzmann equation or
density functional theory \cite{mb97,domps,nc08}.  %Moderately strongly correlated electron systems have been also
%successfully treated by hydrodynamic approaches for certain parameter values \cite{aa11}.

The hydrodynamic
equations describing the 1D
electron gas can then be written as
%1
%-----------------------------------------------------
 \bee
 \fpar{n}{t} + \fpar{}{x}(n u)=0,
 \label{con}
 \ene
%2
%-------------------------------------------------------
\bee
m n\left(\fpar{}{t}+ u \fpar{}{x}\right)u=-e n E
-\fpar{p}{x}
+\frac{\hbar^2 n}{2m} \fpar{}{x} \left(\frac{1}{\sqrt{n}}
\npar{2}{}{x} \sqrt{n}\right),  \label{m1} \ene
 and
%3
%---------------------------------------------------------
 \bee
 \fpar{E}{x}= 4 \pi e (n_0-n),\;\;\;\; \fpar{E}{t}= 4 \pi e n u,
 \label{p1}
\ene
 where $n, u, m$ and $-e$ are resp. the electron number density, fluid velocity, mass and charge, while
$E$ is the electric field, $\hbar$ is the reduced Planck's constant
and $n_0$ is the bulk ionic density background. The RHS of Eq.(\ref{m1}) contains the
quantum statistical pressure  $p=m v_F^2
n^3/(3n_0^2)$, where $m v_{F}^2/2 = \kappa_B T_F = \hbar^2 (3\pi^2 n_{0})^{2/3}/(2m)$ is the Fermi energy, with $\kappa_B$ being Boltzmann's constant. For instance \cite{maha}
with $n\sim 10^{28}m^{-3}$ we get
$T_F\sim 5\times 10^4 K$, much larger than the room temperature, justifying the equation of state for a fully degenerate 1D electron gas.
The third term $\sim \hbar^2$ in the RHS of Eq. (\ref{m1}) is the Bohm potential due
to quantum diffraction. For metallic nanostructures the
quantum statistical pressure and the Bohm potential  are of the same
order  \cite{nc08}, so both contributions will be retained. Finally, for simplifying reasons, exchange-correlation as well as non-ideal, collisional effects are left for future work.

Next we present a procedure to derive exact solutions of Eqs. (\ref{con}) -
(\ref{p1}) by using the  Lagrangian fluid variables method.
%\cite{davidson}. 
We will see how this method can
handle the ``difficult" nonlinearities represented by the
convective $u\partial_x u$ and Bohm potential terms in the
momentum equation.  In a different context, Lagrangian variables have been considered \cite{xxx} to the treatment of the wave-breaking problem in quantum plasmas.
 
We first
transform from Eulerian variables $(x,t)$ to Lagrangian
variables $(\zeta, \tau)$, such that $\zeta = x$ at $t=0$, where
$\tau \equiv  t$ and $\zeta \equiv  x - \int_0^\tau d\tau'\;
u(\zeta, \tau'),$ so that $\zeta$ is a function of both $x$ and
$t$.
%, but $\zeta$ and $\tau$ are treated as independent variables.
In terms of the new variables, the material derivative
$\partial/\partial t + u
\partial/\partial x = \partial/\partial \tau$. Thus, from Eq. (\ref{con}) we
obtain  ${n(\zeta, \tau)}/{n(\zeta,0)}=
{\partial \zeta}/{\partial x}$, where $n(\zeta,0)$ represents the
initial ($\tau=0$) electron density. The
fluid equations, in these variables, are
%-----------------------------------------------------------------------------4
\bee
 \fpar{}{\tau}\left(\frac{1}{n}\right)=\frac{1}{n(\zeta,0)} \fpar{u}{\zeta}
\label{w2}
\ene
and
\begin{eqnarray}
\label{w1}
\Bigl(\frac{\partial^2}{\partial\tau^2} &+& \omega_{p}^2\Bigr)u = \frac{\partial}{\partial\tau}\Bigl\{\frac{-2\kappa_B T_F}{mn_{0}^2}\frac{n^2}{n(\zeta,0)}\frac{\partial n}{\partial\zeta} + \\
&+& \frac{\hbar^2}{2m^2}\frac{n}{n(\zeta,0)}\frac{\partial}{\partial\zeta}\Bigl[\frac{1}{\sqrt{n}}\frac{n}{n(\zeta,0)}\frac{\partial}{\partial\zeta}\Bigl(\frac{n}{n(\zeta,0)}\frac{\partial\sqrt{n}}{\partial\zeta}\Bigr)\Bigr]\Bigr\} . \nonumber
\end{eqnarray}
%
%\beea
%\left(\npar{2}{}{\tau}+\omega_p^2\right)u=\fpar{}{\tau}\{- \frac{2 \kappa_B T_F}{m
%
%n_0^2}\frac{n^2}{n(\zeta,0)} \fpar{n}{\zeta}  +\frac{\hbar^2}{2m^2} \frac{n}{n(\zeta,0)}
%
% \nonumber \\  \fpar{}{\zeta} 
%
%\left[\frac{1}{\sqrt{n}} \frac{n}{n(\zeta,0)} \fpar{}{\zeta}\left(\frac{n}{n(\zeta,0)}
%
%\fpar{\sqrt{n}}{\zeta}\right)\right]\}.
%
%\label{w1}
%\enea
%--------------------------------------------------------------------------------5
The coupled nonlinear partial
differential equations Eqs. (\ref{w2}) and (\ref{w1}) are respectively the continuity and momentum equations
in the Lagrangian variables. The electric field has
been eliminated using the Maxwell equations. Here $\omega_p= \sqrt{4 \pi n_0e^2/m}$ represent the electron plasma frequency.

We seek solutions by separation of variables, assuming  $n(\zeta, \tau)=
N(\zeta)/\phi(\tau)$ and $u(\zeta, \tau) = U(\zeta)\psi(\tau)$.
Inserting into Eqs.(\ref{w2}) and (\ref{w1}) and
separating space and time variable equations, we obtain
%----------------------------------------------------------------------6
\bee
 \frac{1}{\psi}\fder{\hat\phi}{\tau} = \fder{U}{\zeta}=C_1,
 \label{s2}
\ene
and
\beea
\left(\fder{\hat (1/\phi^3)}{\tau}\right)^{-1}\left(\nder{2}{}{\tau}+\omega_p^2\right) \psi(\tau)
= \frac{1}{U(\zeta)}\left\{-\frac{2 \kappa_B T_F}{m}\hat N\fder{\hat N}{\zeta}
 \right. \nonumber\\ \left.+ \frac{\hbar^2}{2m^2} \fder{}{\zeta}
\left(\frac{1}{\sqrt{\hat N}} \nder{2}{}{\zeta}\sqrt{\hat N}\right)\right\}=C_2,
 \label{s1}
\enea
 %7
where $n(\zeta,0)= N(\zeta)/\phi(0)$, $u(\zeta,0)= U(\zeta)
\psi(0)$ with $\psi(0)/\phi(0)\neq 0$ and  where $C_1, C_2$
are arbitrary constants. Here $\hat \phi=\phi /\phi(0)$, $\hat
N = N /N(0)$, and $N(0)/\phi(0)=n_0$.  Eliminating $U, \psi$ between Eqs. (\ref{s2}) and (\ref{s1}) gives the ordinary differential equations
%8
\beea
\frac{d^2\hat\phi}{d\tau^2} + \omega_{p}^2 \hat\phi = \frac{C}{\hat\phi^3},
% \fder{}{\tau}\left(\nder{2}{}{\tau}+\omega_p^2\right)\left( \frac{1}{\hat \phi}\right)
%+ C\fder{\hat \phi^3}{\tau},
 \label{px}
  \\
%and
 %9
%\bee
\fder{}{\zeta} \left(\frac{\hbar^2}{2 m^2} \frac{1}{\sqrt{\hat N}} \nder{2}{}{\zeta}\sqrt{\hat N}
 -\frac{\kappa_B T_F}{m} \hat N^2\right)
 = C \zeta. \label{py}
 \enea
where the separation constant appear as $C = C_1 C_2$.
Note that $C$ has the
dimension of $\mbox{time}^{-2}$. For simplicity, in this Letter two further integration constants were set to zero in the derivation. Equation (\ref{px}) is known as Pinney's equation \cite{pinney} and is endemic in nonlinear analysis. In the autonomous case as in the present application, it has well known explicit analytic solutions, for instance
%10
\bee \hat \phi= \left(1-\hat \alpha^2 \sin^2
\omega_p\tau\right)^{1/2}, \label{tsol} \ene where $\hat
\alpha=\alpha/\omega_p.$ We
have used the initial condition $\hat \phi=1$, $d \hat
\phi/d\tau=0$ and $d^2 \hat \phi/d\tau^2=-\alpha^2$, at $\tau = 0$. In other words, a periodic (cosine-like) initial condition with frequency
$\alpha$ have been chosen, where we have taken the separation constant $C=
\omega_p^2-\alpha^2$. The solution (\ref{tsol}) gives the temporal part of the density modulation.
The parameter $\hat \alpha^2$ determines how much energy is
transported from the initial source to the electron wave i.e. it
measures the amplitude of the initial density perturbation. We need $\hat \alpha^2 <1$ or equivalently $C > 0$ for a well behaved solution. In other words, a collapse will be avoided if the RHS of Eq. (\ref{px}) is repulsive. Pinney's equation also arise for other quantum plasma problems, like in the case of the quantum Buneman instability \cite{Bret}.

Next we consider the space part. Initially, notice the first $\sim \hbar^2$ term inside the bracket in the LHS of Eq. (\ref{py})
originate from the kinetic energy term in the Schr\"odinger equation, while the remaining $\sim T_F$ term contains the Fermi pressure contribution.
To rewrite Eq. (\ref{py}), we use
initial conditions $\hat N=1$ and
$(\hbar^2/2m\sqrt{\hat N})d^2\sqrt{\hat
N}/d\zeta^2=-\epsilon$ at $\zeta = 0$. Here $\epsilon$ is the quantum kinetic
energy at $\zeta=0$. Also
the parameter $\epsilon$ serves as a measure of the dispersive effects, since
it originates from the quantum diffraction term. Integrating once, the space part
equation now becomes
 %11
 \bee
-\frac{\hbar^2}{2m }
\nder{2}{\chi}{\zeta}+\left(\frac{1}{2} m a^2 \omega_p^2
\zeta^2 - \varepsilon\right)\chi + \kappa_B T_F \chi^5=0,
\label{qho}
\ene
 where $\chi=\sqrt{\hat N},\; a=\sqrt{1-\hat \alpha^2}$ and
$\varepsilon=\epsilon + \kappa_B T_F,$. The nonlinear-Schr\"odinger-like equation (\ref{qho}) describe nonlinear oscillations in Lagrangian variables where the anharmonicity is due to the Fermi temperature $T_F \neq 0$. Similar nonlinear
equations arise in related areas, e.g. in the analysis of low-dimensional Bose-Einstein condensates confined by harmonic traps \cite{ek00}.

In the Thomas-Fermi
approximation, the quantum (Bohm) potential is neglected, which
amounts to neglecting the derivative and $\sim \epsilon$ terms in
Eq. (\ref{qho}). In this limit, the density profile becomes
%12
\bee
 n_{TF}(x, t)=\frac{n_0}{\sqrt{1-\hat \alpha^2 \sin^2 \omega_p \tau}}
\left(1- \frac{m a^2 \omega_p^2 \zeta^2}{2\kappa_B T_F}\right)^{1/2},
 \label{tf}
 \ene
in the region $m a^2 \omega_p^2
\zeta^2 \leq 2\kappa_B T_F$, with the density being zero outside this region. 

\begin{figure}[h]
\centering
%\begin{tabular}{cc}
{\includegraphics[width=2.0in,height=1.6in]{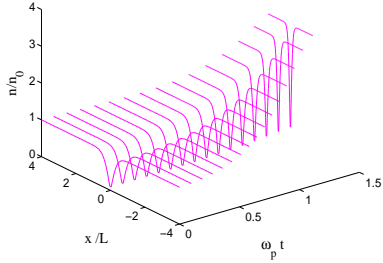}}
%& {\includegraphics[width=1.7in,height=1.2in]{f5.eps}}
 %\end{tabular}
\caption{Normalized density evolution according to Eq. (\ref{bec}), with
  $\hat\alpha^2=0.9$ and $\hbar\omega_p/\kappa_B T_F=0.2$. The length sale
 is normalized by $L=(\hbar/m \omega_p)^{1/2}.$}
 \label{x0}
 \end{figure}

For sufficiently large initial perturbation or $(\hat\alpha^2 \rightarrow
1)$, the term containing $a$ in Eq. (\ref{qho}) can be omitted. 
Then for negligible quantum kinetic energy compared to internal
energy i.e. for $\epsilon\ll \kappa_B T_F$, for certain carefully chosen initial conditions we obtain the following
approximate solution of Eq. (\ref{qho}), 
%13
\bee
 n(x, t)=\frac{n_0}{\sqrt{1-\hat \alpha^2 \sin^2 \omega_p \tau}}
 \left(\frac{\cosh 2 \bar \zeta -1} {\cosh 2 \bar \zeta +2}\right),
 \label{bec}
 \ene
 where $\bar \zeta= (2m\kappa_B T_F/\hbar^2)^{1/2}\zeta$.
It is interesting to note that the spatial density profile
is like a gray soliton as shown in Fig.  \ref{x0}. The neglect of the restoring harmonic $\sim \zeta^2$ term is, however, valid for a short interval such that 
$m a^2 \omega_{p}^2 \zeta^2 \ll \varepsilon$.

\begin{figure}[h]
\centering
%\begin{tabular}{cc}
{\includegraphics[width=2.0in,height=1.6in]{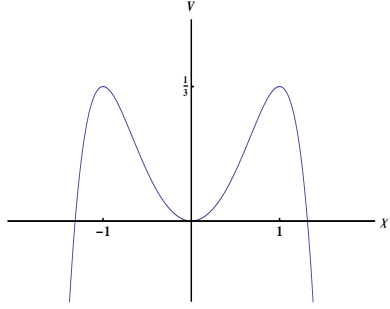}}
%& {\includegraphics[width=1.7in,height=1.2in]{f5.eps}}
 %\end{tabular}
\caption{Pseudo-potential from Eq. (\ref{pp}), with $\varepsilon = \kappa_B T_F$ so that $V_{\rm max} = 1/3$.}
 \label{xx}
 \end{figure}

It is instructive to examine more closely the exactly resonant case $\hat\alpha = 1, a = 0$, for which $\hat\phi = \cos\omega_p \tau$ and Eq. (\ref{qho}) reduces to 
\bee
\label{res}
\frac{\hbar^2}{2m}\nder{2}{\chi}{\zeta} + \varepsilon\chi - \kappa_B T_F \chi^5=0,
\ene
which is the equation for an autonomous quintic oscillator. The equation of motion for $\chi$ is reducible to quadrature and the exact solution can be expressed in terms of elliptic functions. 
Indeed, we may rewrite 
\bee
\label{ress}
\frac{d^2\chi}{d\bar\zeta^2} = - \frac{dV}{d\chi} \,,
\ene
with the pseudo-potential 
\bee
\label{pp}
V = V(\chi) = \frac{\varepsilon}{\kappa_B T_F}\frac{\chi^2}{2} - \frac{\chi^6}{6} \,.
\ene
From the form of $V(\chi)$ (see Fig. (\ref{xx})), it follows that periodic motion in the space-like variable $\zeta$ appear when the energy integral $(d\chi/d\bar\zeta)^2/2 + V(\chi) < V_{\rm max}$, where the maximum value $V_{\rm max} = (1/3)[\varepsilon/(\kappa_B T_F)]^{3/2}$, and suitable initial value $\chi(0)$ between the turning points. The energy integral can be also used to produce the aforementioned quadrature. 
A simple analysis of the energy conservation law associated to (\ref{ress}) shows that necessary conditions for nonlinear bounded oscillations are given by $\chi^{2}(0) < 
[\varepsilon/(\kappa_B T_F)]^{1/2}$ and $[(d\chi/d\bar\zeta)(0)]^2 < (2/3)[\varepsilon/(\kappa_B T_F)]^{3/2}$, which is interpreted as follows. The quantity $\chi(\bar\zeta)$ and the derivative are related to density perturbations and can not be too large, otherwise the diffusive effects due to the Bohm potential and the degeneracy pressure will predominate. On the other hand, a sufficient high $\varepsilon$ (or, a sufficiently high restoring electric force related to the plasma frequency $\omega_p$) tends to enlarge the phase space region for which oscillatory motion is possible. Notice, however, that in the exact resonance case treated here the solution will develop a singularity in a finite time since $\hat\phi \rightarrow 0$ as $\tau \rightarrow (\pi/2)\omega_{p}^{-1}$. In this context, the localized solution (\ref{bec}) can be shown to correspond to separatrix motion, in the border between periodic and aperiodic trajectories.

In yet another case, when the internal energy is negligible in
comparison with the quantum kinetic energy i.e. $\kappa_B T_F \ll
\epsilon $, Eq. (\ref{qho}) becomes formally identical to the
time-independent Schr\"odinger equation for the simple
harmonic oscillator, in Lagrangian coordinates. The solution is in terms of the Hermite polynomials $H_\nu$, with
eigenvalues $\epsilon = \epsilon_\nu= a \hbar \omega_p \left(\nu +\frac{1}{2}\right)$,  where $\nu = 0, 1, 2,...$.
More explicitly, disregarding more involved forms involving linear combinations of the elementary solutions, in the limit of  negligible Fermi temperature we get
%--------------------------------------------------------14
\beea
 n(x, t)=\frac{n_0}{\sqrt{1-\hat \alpha^2 \sin^2 \omega_p \tau}}
 H_\nu^2 \left[\left(\frac {a m \omega_p}{\hbar}\right)^{1/2} \zeta \right]
 \nonumber\\
 \exp\left({-\frac {a m \omega_p}{\hbar}\zeta^2}\right).
 \label{hos}
 \enea

\begin{figure}[h]
\centering
%\begin{tabular}{cc}
{\includegraphics[width=2.0in,height=1.6in]{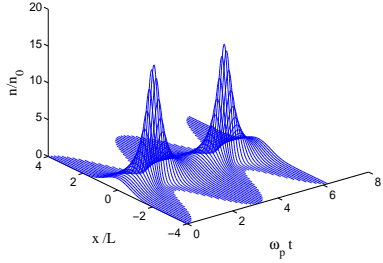}}
%& {\includegraphics[width=1.7in,height=1.2in]{f5.eps}}
 %\end{tabular}
\caption{Normalized density evolution according to Eq. (\ref{hos}), with
  $\hat\alpha^2=0.9$ and $\nu=0$. The figure shows  that for
  $\hat \alpha^2 \rightarrow 1$ the density goes towards
 quasi-singularity in the absence of Fermi temperature. The length sale
 is normalized by $L=(\hbar/a m \omega_p)^{1/2}.$}
 \label{x1}
 \end{figure}

 Figure (\ref{x1}) show the space-time behavior of the density
$n$ given by (\ref{hos}) for  parameters $\hat \alpha^2 =0.9$ and
$\nu  =0$. A quasi-collapse is seen by the blowing up
of $n$  just before $ t_c=(\pi/2) \omega_{p}^{-1}$. The spatial width of the density appear as a pulsating breather form after transforming back to Eulerian variables. Multi-breather solutions can be obtained for $\nu = 1, 2, 3...$

In the above approximate solutions, the relation
between $\zeta$ and $x$ is $ x = \zeta
{(1-\hat\alpha^2 \sin^2 \omega_p \tau)^{1/2}}.$ Furthermore, the fluid
electron velocity $u(\zeta,\tau) = -\hat \alpha^2 \omega_p \zeta \sin \omega_p \tau
\cos \omega_p \tau/{(1-\hat \alpha^2 \sin^2 \omega_p
\tau)^{1/2}}.$  Equations represent in limit cases the
solution of Eq. (\ref{qho}) in three
different physical situations. The parameter $\hat \alpha^2$ stands for the strength
of the nonlinearity and controls the amplitude of the density modulations. At
the time $\tau=(\pi/2) \omega_{p}^{-1}$, one finds that for $\hat\alpha^2
\rightarrow 1$ there appears a quasi-density singularity. Actually
$\hat\alpha \sim 1$ gives the resonance condition $\alpha\sim
\omega_p$. At this resonance, the initial periodic source pumps the
maximum energy into the system, resulting in an inward cavitation blowing up the density and shrinking the inhomogeneity
width.

\begin{figure}
\centering
\begin{tabular}{cc}
{\includegraphics[width=1.7in,height=1.2in]{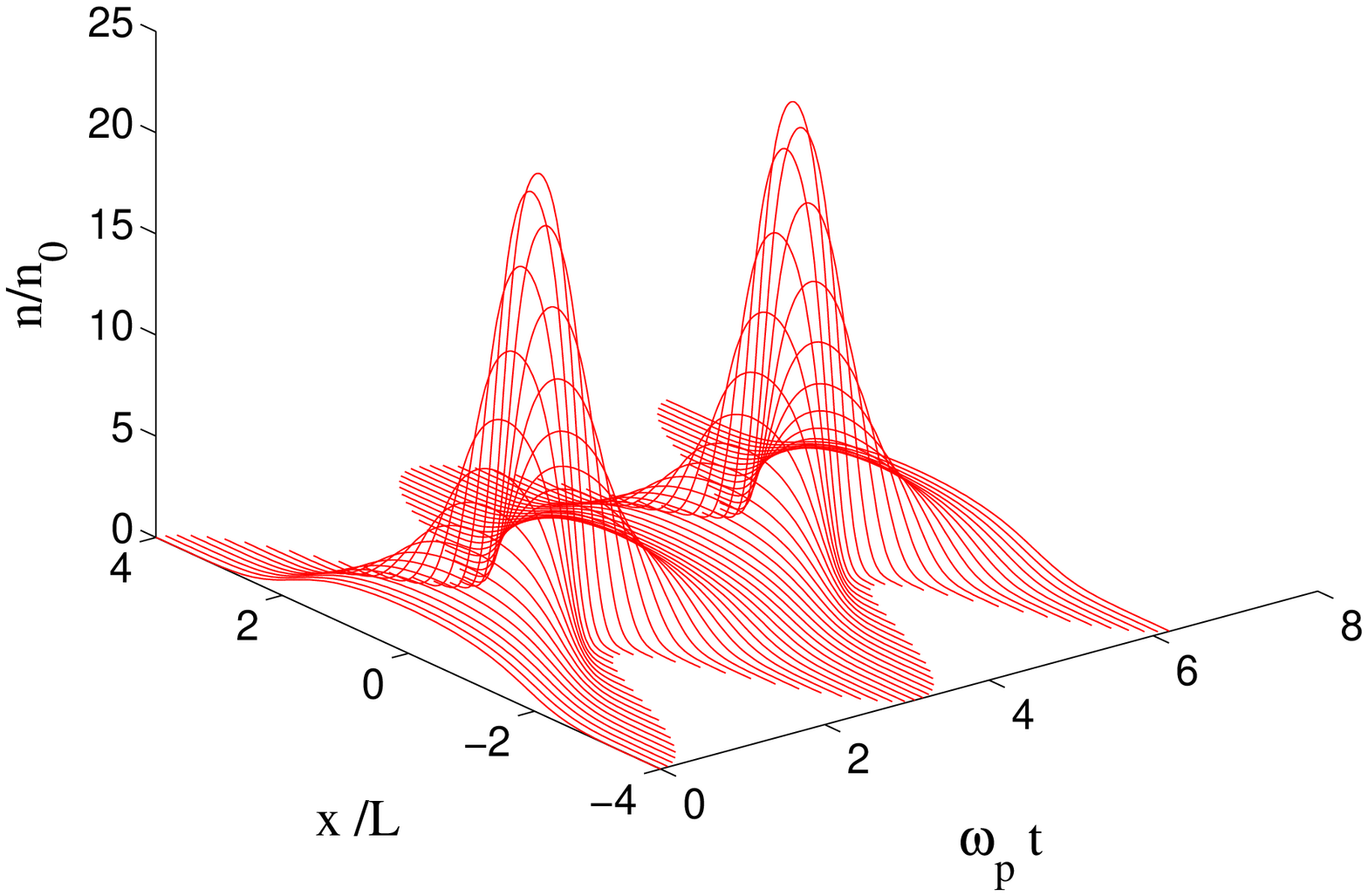}}
& {\includegraphics[width=1.7in,height=1.2in]{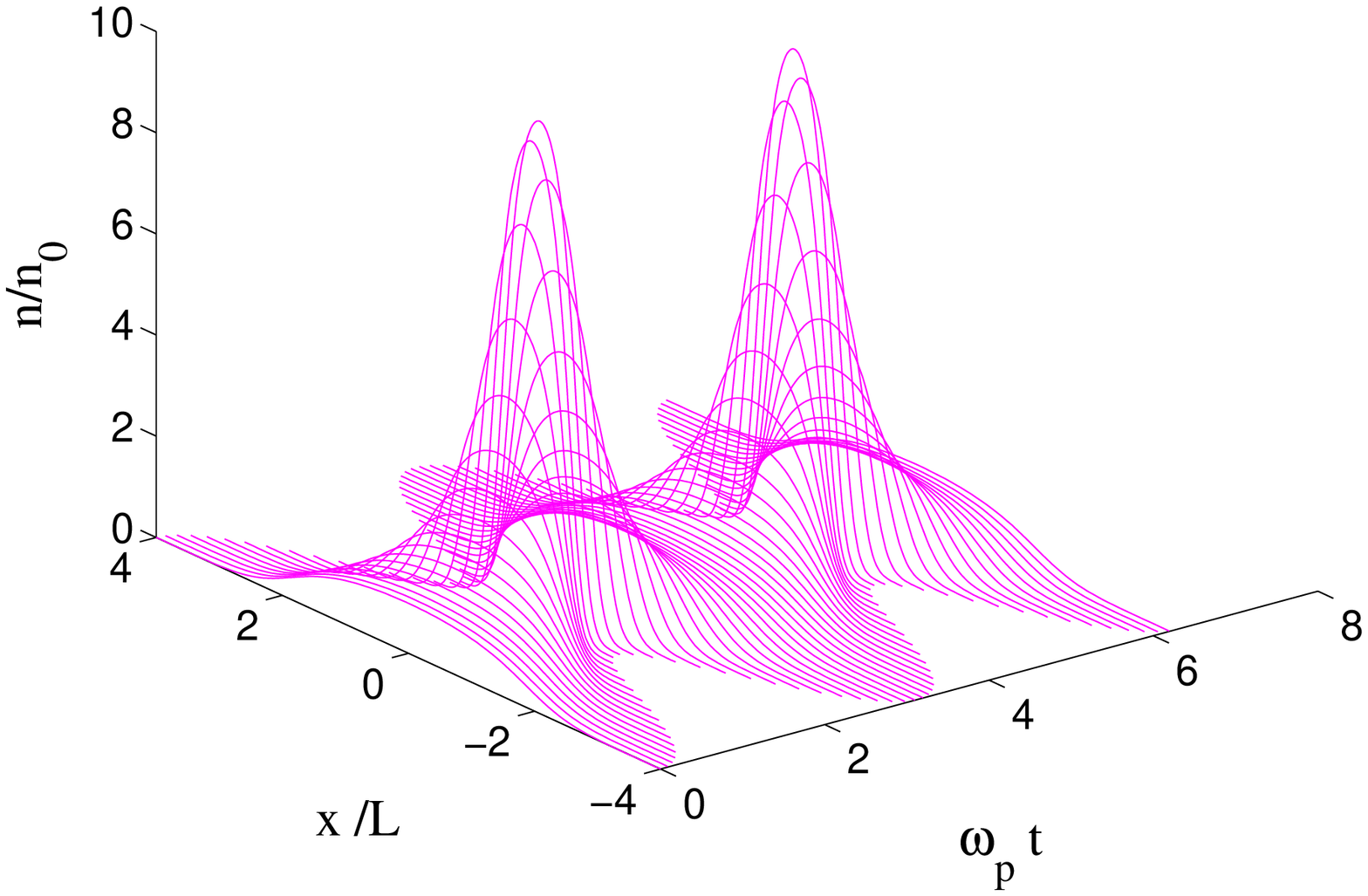}}
 \end{tabular}
\caption{Normalized density evolution with
$\hat\alpha^2=0.9$. The left-hand side figure shows that for
$\varepsilon/a\hbar \omega_p=5.0$ and $\kappa_B T_F/a\hbar \omega_p=1.0$ the 
density tends to a singularity. Increasing the value of  $\kappa_B
T_F/a\hbar \omega_p=5.0$, the right-hand side figure exhibits the reduction of the density amplitude and spatial dispersion due to the degeneracy pressure. The spatial variable $x$  is normalized by
$L=(\hbar/a m \omega_p)^{1/2}.$}
 \label{x2}
 \end{figure}

\begin{figure}
\centering
\begin{tabular}{cc}
 {\includegraphics[width=1.60in,height=1.2in]{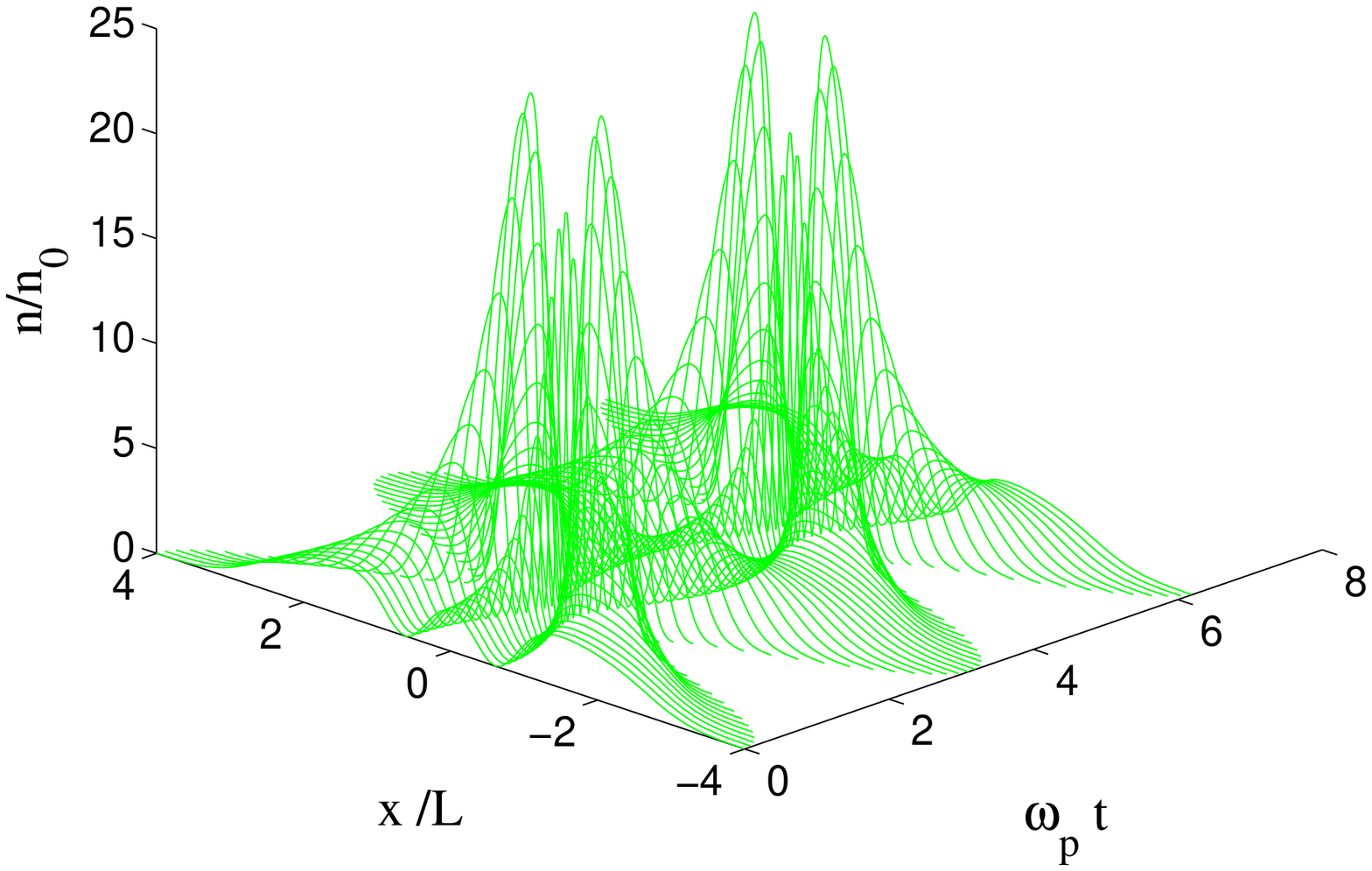}}
 & {\includegraphics[width=1.60in,height=1.2in]{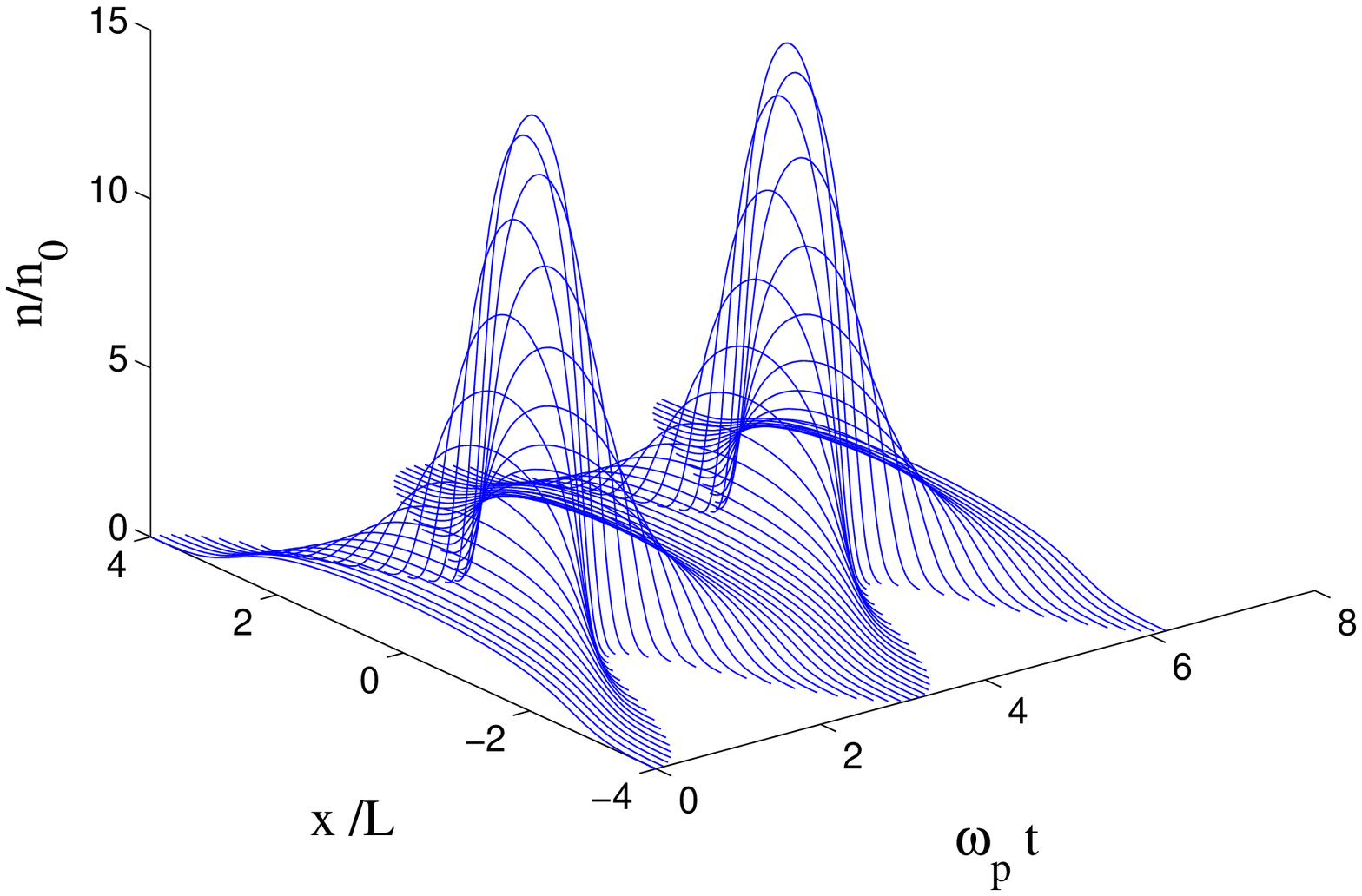}}
 \end{tabular}
\caption{The left side figure shows the
electronic density  for $\varepsilon/a\hbar \omega_p=9.0$, $\kappa_B
T_F/a\hbar \omega_p=1.0$. A multi-breather structure in space and time appear for a small electron Fermi 
temperature. In the right, a bigger Fermi temperature so that $\kappa_B
T_F/a\hbar \omega_p=4.0$ produce a single breather solution. The length scale is normalized by $L=(\hbar/a m
\omega_p)^{1/2}.$} \label{x3}
 \end{figure}

Finally, we numerically simulate Eqs. (\ref{tsol}) and
(\ref{qho}) using a finite difference scheme. The
density profile is shown in Fig. (\ref{x2}) for $\hat
\alpha^2=0.9$, total energy parameter $\varepsilon/a\hbar
\omega_p=5.0$ and internal energy parameter $\kappa_B T_F/a\hbar
\omega_p=1.0$. Close to collapse the density
becomes strongly peaked and narrow, remaining finite.
Increasing the internal energy up to $\kappa_B T_F/a\hbar \omega_p=5.0$
substantially reduces the density amplitude and
finally stable breather structures are formed as shown in Fig.
(\ref{x2}). The reduction in density amplitude is attributed to
the dispersive effect caused by the quantum statistical pressure.
For larger value of the total energy parameter $\varepsilon/a\hbar
\omega_p=9.0$ with $\kappa_B T_F/a\hbar \omega_p=1.0$, a multi-breather structure for the density is developed in space. However, for
higher Fermi temperature (or, higher internal energy) $\kappa_B T_F/a\hbar
\omega_p=4.0$, the multi-breather structure reduces to a
single breather, which is depicted in Fig. (\ref{x3}). This is because the restoring harmonic term $\sim \zeta^2$ in Eq. (\ref{qho}), which produces oscillations arising from the initial density perturbations, is overcome by the nonlinearity due to the degeneracy pressure.
 %------------------

In conclusion, we have demonstrated a new route to generate nonlinear electron waves in a fully degenerated 1D electron gas, using
Lagrangian variables. The nonlinear structures are described by
ordinary differential equations, in significant reduction in comparison to the
spatio-temporal fluid equations. Altough the focus of the work was on the density oscillations, similar conclusions can be derived for the electric field and quantum fluid velocity. 
%The anharmonicity is due to the Fermi
%electron pressure and the oscillator kinetic energy which
%serves as the source of dispersion arising from the quantum potential
%associated with the electron tunneling effect (through the Bohm term). %The competition between multiple nonlinearities (the electron advection, the quantum statistical pressure, and the nonlinear quantum Bohm
%potential) and the dispersion due to the quantum statistical
%pressure lead to localized
%soliton-like density structures.
In the limit of large amplitude initial density perturbations (i.e. nearly the resonance $\hat \alpha^2
\rightarrow 1$), the plasma density develop a quasi-singular cavitation in finite time. However, the catastrophic collapse can be avoided for specific
parameters allowing the formation of stable breather structures, thanks to the Fermi pressure and Bohm potential contributions.
With respect to the stability, the robustness of the new analytical solutions has been numerically verified. On one hand the new stable, long-lived coherent nonlinear structures can
be useful for information transport at nanoscales. On the other hand, the derived analytical estimates are useful tools for
the determination of the necessary parameters to avoid collapse in real systems. As a general rule, the use of Lagrangian variables in the
context of quantum fluid equations is a very promising new avenue for quantum hydrodynamic equations arising in different fields,
like the metallic nanostructures considered here, quantum plasmas \cite{maha2}, graphene, plasmonics, or quantum diodes \cite{eliasson}, whenever quantum fluid equations are applicable. Further effects, such as dissipation and exchange-correlation energies can significantly change the behaviors found here, but are outside the scope of the present work. 

%% here a revision

%\revision{Insert here the text.
%See fig.~\ref{fig.1}, table~\ref{tab.1} and eq.~(\ref{eq.1}).
%See also~\cite{b.a,b.b}.}

%here a shortcut $\emc$ and again $\emc$

%\begin{equation}
%\label{eq.1}
%0\neq1
%\end{equation}

%\begin{figure}
%\onefigure{epl-template.eps}
%\caption{Figure caption.}
%\label{fig.1}
%\end{figure}

%\begin{table}
%\caption{Table caption.}
%\label{tab.1}
%\begin{center}
%\begin{tabular}{lcr}
%first  & table & row\\
%second & table & row
%\end{tabular}
%\end{center}
%\end{table}

\acknowledgments
One of us (FH) thanks the Brazilian agency CNPq for financial support.

\end{document}